\documentstyle[preprint,aps]{revtex}
\def\no{\nonumber}
\def\a{\alpha}
\def\d{\delta}
\def\e{\epsilon}
\def\m{\mu}
\def\n{\nu}
\def\LLL{\Lambda}
\def\lll{\lambda}
\def\mn{{\mu\nu}}

\def\p{\phi}
\def\eq{\equiv}
\def\pa{\partial}
\def\ra{\rightarrow}
\def\lag{{\cal L}}
\newcommand{\np}[1]{Nucl.\ Phys.\ {\bf {#1}}}
\newcommand{\plt}[1]{Phys.\ Lett.\ {\bf {#1}}}
\newcommand{\pr}[1]{Phys.\ Rev.\ {\bf {#1}}}
\newcommand{\prlt}[1]{Phys.\ Rev.\ Lett.\ {\bf {#1}}}

\newcommand{\fumn}{F^{\m\n}}
\newcommand{\humn}{H^{\m\n}}
\newcommand{\fdmn}{F_{\m\n}}
\newcommand{\hdmn}{H_{\m\n}}
\newcommand{\dbm}{\bar{D}_\m}
\newcommand{\dbn}{\bar{D}_\n}
\newcommand{\calf}{{\cal F}}
\newcommand{\calh}{{\cal H}}

\begin{document}
\preprint{SNUTP 95-035\raisebox{-15pt}{\hspace{-27.5mm}CU-TP-681}}
%\draft
\title{The Bogomol'nyi Bound of Lee-Weinberg\\
Magnetic Monopoles}
\author{Choonkyu Lee}
\address{Department of Physics and Center for Theoretical Physics\\
Seoul National University, Seoul, 151-742, Korea}
\author{and}
\author{Piljin Yi\thanks{e-mail address: piljin@cuphyg.phys.columbia.edu}}
\address{Physics Department, Columbia University\\ New York, NY 10027, U.S.A}
\maketitle
\begin{abstract}
The Lee-Weinberg $U(1)$ magnetic monopoles, which have been reinterpreted as
topological solitons of a certain non-Abelian gauged Higgs model recently, are
considered for some specific choice of Higgs couplings. The model under
consideration is shown to admit a Bogomol'nyi-type bound which is saturated by
the configurations satisfying the generalized BPS equations. We consider the
spherically symmetric monopole solutions in some detail.
\end{abstract}
\newpage
Recently, Lee and Weinberg \cite{1} constructed a new class of finite-energy
magnetic monopoles in the context of a purely Abelian gauge theory. The
corresponding $U(1)$ gauge potential is simply that of a point-like Dirac
monopole \cite{2} (with the monopole strength satisfying the Dirac
quantization condition), yet the total energy is rendered finite by
introducing a charged vector field of arbitrary positive gyromagnetic ratio
and by fine-tuning a quartic self-interaction. This is in a marked contrast to
finite-mass magnetic monopoles of the 't Hooft-Polyakov type \cite{3}, which
appear as topological solitons of some spontaneously broken non-Abelian gauge
theories. In the latter case, the existence of such solitons and the charge
quantization \cite{2} thereof are understood in terms of the nontrivial second
homotopy group of the appropriate vacuum manifold.

In Ref.\ \cite{4}, however, we realized that the above Lee-Weinberg monopole
is also equipped with a hidden, spontaneously broken $SO(3)$ gauge symmetry.
This naturally explains the integer-charged Lee-Weinberg monopoles as
topological solitons associated with the vacuum manifold $SO(3)/U(1)$, just as
in the usual $SO(3)$ Higgs model. In the present paper, we will determine the
{\it self-dual} limit of the Lee-Weinberg model, in which the energy
functional satisfies a simple Bogomol'nyi-type bound. This is achieved
for a special form of the Higgs potential, but the gyromagnetic ratio
can still assume an arbitrary positive value. [For $g=2$, our model
reduces to the old Bogomol'nyi-Prasad-Sommerfield (BPS) model \cite{5}.]
Configurations that saturate the thus-obtained Bogomol'nyi bound solve
certain first-order differential equations; they generalize the old BPS
equations, the investigation of which has been an important part of
mathematical physics for
the last two decades \cite{6}. These generalized BPS monopoles satisfy the
same mass formula as the old BPS monopoles. Also given is a simple argument
which demonstrates the existence of unit-charged monopole solutions to our
generalized BPS equations, while satisfying the physical boundary conditions.
There is a strong evidence that static multi-monopole solutions exist in this
model as well.

Let us recapitulate the observations of Refs.\ \cite{1} and \cite{4} first.
The Abelian model of Ref.\ \cite{1} consists of a $U(1)$ electromagnetic
potential $A_\m$, a charged vector field $W_\m$, and a real scalar $\p$, with
the Lagrangian density chosen to have the form
\begin{eqnarray} \label{e1}
\lag&=&-\frac{1}{4}\fumn\fdmn-\frac{1}2|\dbm W_\n-\dbn W_\m|^2
       +\frac{g}{4}\hdmn\fumn-\frac{\lll}{4}\hdmn\humn\no\\
    &&-\frac{1}2\pa_\m\p\pa^\m\p-m^2(\p)|W_\m|^2-V(\p)\,,
\end{eqnarray}
where $\fdmn\eq\pa_\m A_\n-\pa_\n A_\m$, $\dbm W_\n\eq\pa_\m W_\n+ieA_\m W_\n$,
and $\hdmn\eq ie(W_\m^*W_\n-W_\n^*W_\m$). The coupling $g$ is the gyromagnetic
ratio associated with the magnetic moment of the charged vector, and $m^2(\p)$
is assumed to vanish at $\p=0$ but is equal to $m_W^2(\neq0)$ when $\p$ is at
its (nontrivial) vacuum value. With $g=2$, $\lll=1$ and $m(\p)=e\p$, this is
nothing but the unitary gauge version of the spontaneously broken non-Abelian
gauge theory of Ref.\ \cite{3} and thus renormalizable; but for generic values
of $g$ and $\lll$, the theory is nonrenormalizable. Author of Ref.\ \cite{1}
noted that if the couplings satisfy the relation $\lll=\frac{g^2}{4}$ (for
arbitrary positive $g$), the model allows magnetic monopoles with {\it
finite} mass.

The above model can be recast as a nonrenormalizable $SO(3)$ gauge theory with
the gauge connection 1-form $B=(B_\m^adx^\m)T^a$ and a triplet Higgs
$\Phi=\Phi^aT^a$. Specifically, consider the theory defined by the Lagrangian
density \cite{4}
\begin{eqnarray} \label{e2}
\lag'&=&-\frac{1}4G_{\m\n}^aG^{a\m\n}+\frac{g-2}4\calf_{\m\n}\calh^{\m\n}
        -\frac{\lll-1}4\calh_{\m\n}\calh^\mn
        -\frac{1}2(D_\m\Phi)^a(D^\m\Phi)^a\no\\
     &&-\frac{1}{e^2}(m^2(|\p|)-e^2\Phi^a\Phi^a)
       (D_\m\hat\Phi)^a(D^\m\hat\Phi)^a
       -V(|\Phi|)\,,
\end{eqnarray}
where $G$ is the non-Abelian field strength associated with $B$,
$(D_\m\Phi)^a\eq\pa_\m\Phi^a+e\e^{abc}B_\m^b\Phi^c$,
$\hat\Phi^a\eq\Phi^a/|\Phi|$, and we have defined two (gauge-invariant)
tensors $\calf_\mn$, $\calh_\mn$ by
\begin{equation}
\calf_\mn-\calh_\mn=G_\mn^a\Phi^a\,,\hspace{5mm}
\calh_\mn=-\frac{1}e\e_{abc}\hat\Phi^a(D_\m\hat\Phi)^b(D_\n\hat\Phi)^c\,.
\end{equation}
Note that 't Hooft \cite{3} previously used the tensor $\calf$ to represent
physical electromagnetic fields. Now, in the unitary gauge (i.e.,
$\hat\Phi^a=\d^{a3})$, we may identify $|\Phi|$ with $|\p|$, $B_\m^3$ with
$A_\m$, and $\frac{1}{\sqrt{2}}(B_\m^1+iB_\m^2)$ with $W_\m$; then, we have
$\calf_\mn=\fdmn$, $\calh_\mn=\hdmn$,
$(D^\m\hat\Phi)^a(D_\m\hat\Phi)^a=2e^2W_\m^* W^\m$, etc. As a result, the
$SO(3)$ invariant Lagrangian density $\lag'$ reduces to the expression
(\ref{e1}) of the apparently Abelian theory of Lee and Weinberg. This in turn
allows us to reinterpret all integer-charged Lee-Weinberg monopoles as
topological solitons of the non-Abelian theory defined by $\lag'$.
In the latter
description we can have the monopoles described in a non-singular (i.e.,
string-free) gauge, and if $n$ is the winding number associated with the map
$\hat\Phi^a(r=\infty):S^2\ra S^2$, the magnetic monopole strength is simply
given by $g_{\text{magnetic}}=-\frac{4\pi n}{e}$ \cite{7}. Especially, the
unit-charged $(n=\pm1)$ Lee-Weinberg monopoles may be described by the
familiar hedgehog form:
\begin{equation} \label{e4}
\Phi^a=\hat x^a\p(r)\,,\hspace{5mm}
B^a=-\frac{\e_{abc}x^bdx^c}{er^2}[1-u(r)]\,.
\end{equation}

Aside from the topological argument, one generally needs to look into the
energetics at short distances to ascertain the existence of actual
finite-energy monopole solutions. As mentioned already, Lee and Weinberg
showed that the monopoles have finite energy if $\lll=g^2/4$ with arbitrary
positive $g$; this conclusion was confirmed also \cite{4} using the
spherically symmetric form (\ref{e4}) in the equivalent non-Abelian
description. But, $u^2(0)=\frac{2}g$ for these Lee-Weinberg monopoles and thus
the corresponding vector potentials are not regular at the origin (except for
the special case $g=2$, corresponding to the renormalizable model). More
detailed study on these solutions may be carried out with the help of the
field equations (with $\frac{g^2}4=\lll$). For the radial functions $\p(r)$
and $u(r)$, they imply
\begin{mathletters}
\begin{eqnarray}
&&\frac{d^2u}{dr^2}+\frac{g}2\left(1-\frac{g}2u^2\right)\frac{u}{r^2}
         =m^2(\p)u\,,\\
&&\frac{d^2\p}{dr^2}+\frac{2}r\frac{d\p}{dr}
    -\frac{u^2}{e^2r^2}\frac{dm^2(\p)}{d\p}=\frac{dV(\p)}{d\p}\,.
\end{eqnarray}
\end{mathletters}%
Making a local analysis with these equations near the origin, it is then easy
to show that $u(r)$ has the following behavior near the origin:
\begin{equation}\label{e6}
\left[u^2(r)-\frac{2}g\right]\sim r^\a\,,\hspace{5mm}
\mbox{with }\a=\frac{1}2+\sqrt{g+\frac{1}4}\,.
\end{equation}
Note that if the gyromagnetic ratio takes the special value $g=2$, this
reduces to the expected analytic behavior with $\a=2$. The behavior
(\ref{e6}) will be sufficient to remove possible boundary contribution at the
origin, thus enabling us to find the Bogomol'nyi bound through the usual
argument.

We may also recall the usual BPS limit of the renormalizable case, which is
realized when the Higgs potential $V(\p)$ is dropped and the couplings are
such that $1=\frac{g}2=\lll=m^2(\p)/e^2\p^2$. The energy functional for purely
magnetic, static configurations can then be expressed as \cite{5}
\begin{eqnarray} \label{e7}
{\cal E}_0&=&\int d^3x\left\{\frac{1}4G_{ij}^aG_{ij}^a
              +\frac{1}2(D_i\Phi)^a(D_i\Phi)^a\right\}\no\\
 &=&\int d^3x\frac{1}2\left\{(D_i\Phi)^a\mp\frac{1}2\e_{ijk}G_{jk}^a\right\}
    \left\{(D_i\Phi)^a\mp\frac{1}2\e_{ijk}G_{jk}^a\right\}\no\\
 &&\pm\int d^3x\frac{1}2\e^{ijk}\pa_i(\Phi^aG_{jk}^a)\no\\
 &\ge&\left|\frac{1}2\oint_{r=\infty}dS_i\e^{ijk}\Phi^aG_{jk}^a\right|\,,
\end{eqnarray}
where we have used the non-Abelian Bianchi identity in performing the partial
integration. The surface integral is equal to $-\frac{4\pi n}{e}\p(\infty)$,
and thus we are led to the Bogomol'nyi bound
\begin{equation}
{\cal E}_0\ge\left|\frac{4\pi n}{e}\p(\infty)\right|.
\end{equation}
The so-called BPS monopoles (or self-dual monopoles) are configurations that
saturate this energy bound. It follows from the bulk part of (\ref{e7}) that
they must solve the BPS equations (or self-duality equations)
\begin{equation} \label{e9}
(D_i\Phi)^a=\pm\frac{1}2\e_{ijk}G_{jk}^a\,,
\end{equation}
where the sign is determined by that of $n$. Especially, if we insert the
hedgehog ansatz (\ref{e4}) into (\ref{e9}), we obtain the first-order
equations for the unit-charged BPS monopoles:
\begin{equation} \label{e10}
\frac{du}{dr}=\pm e\p u\,,\hspace{5mm}
e\frac{d\p}{dr}=\mp\frac{1}{r^2}(1-u^2)\,.
\end{equation}
The solutions to these equations are given in terms of elementary functions.
[See (\ref{e21}) below.]

Now the question is whether we can have a similar Bogomol'nyi limit for
general values of the gyromagnetic ratio $g$ within the model defined by the
Lagrangian density (\ref{e2}). The first clue that this might be possible
comes from studying the energy functional of the given model for spherically
symmetric configurations (i.e., for the form (\ref{e4})):
\begin{equation} \label{e11}
{\cal E}=\int d^3x\left\{\frac{1}{e^2r^2}\left(\frac{du}{dr}\right)^2
       +\frac{1}2\left(\frac{d\p}{dr}\right)^2
       +\frac{1}{2e^2r^4}[\lll u^4-gu^2+1]
       +\frac{u^2}{e^2r^2}m^2(\p)+V(\p)\right\}\,.
\end{equation}
We again drop the Higgs potential $V(\p)$, but keep the gyromagnetic ratio
$g(>0)$ arbitrary such that $\frac{g^2}{4}=\lll$ and $m^2(\p)=\lll e^2\p^2$.
Recall that $\frac{g^2}4=\lll$ was necessary to ensure the finite total
energy. Then, the energy functional in (\ref{e11}) may be rewritten as
\begin{eqnarray} \label{e12}
{\cal E}&=&4\pi\int_0^\infty dr\left\{\left(\frac{1}e\frac{du}{dr}
    \mp\frac{g}2u\p\right)^2
    +\frac{1}2\left(r\frac{d\p}{dr}
     \pm\frac{1}{er}\left(1-\frac{g}2u^2\right)\right)^2\right\}\no\\
&&\mp\frac{4\pi}e\int_0^\infty dr\frac{d}{dr}\left[\p(1-\frac{g}2u^2)\right].
\end{eqnarray}
As long as the combination $\p(1-\frac{g}2u^2)$ vanishes at the origin (which
is true for our case), ${\cal E}$ is manifestly bounded  below by the value
$|4\pi\p(\infty)/e|$. The generalized first-order equations, which can be read
off from the bulk part of (\ref{e12}), are
\begin{equation} \label{e13}
\frac{du}{dr}=\pm\frac{g}2e\p u\,,\hspace{5mm}
e\frac{d\p}{dr}=\mp\frac{1}{r^2}\left(1-\frac{g}2u^2\right).
\end{equation}
It is comforting to see that, in the renormalizable limit $\frac{g}2\ra 1$,
these reduce to (\ref{e10}).

We will now demonstrate that once we choose $V(\p)=0$, $g^2/4=\lll$ and
$m^2(\p)=\lll e^2\p^2$ with the Lagrangian density (\ref{e2}), a generalized
Bogomol'nyi system results without the assumption of the spherical symmetry.
The static energy functional for purely magnetic configurations (i.e., with
$B_0^a\eq 0$) reads
\begin{eqnarray}
{\cal E}&=&\int d^3x\left\{\frac{1}4\left(G_{ij}^a
           -(\frac{g}2-1)\hat\Phi^a\calh_{ij}\right)
           \left(G_{ij}^a-(\frac{g}2-1)\hat\Phi^a\calh_{ij}\right)\right.\no\\
        &&\hspace{12mm}\left.+\frac{1}2
         \left((D_i\Phi)^a+(\frac{g}2-1)\p(D_i\hat\Phi)^a\right)
         \left((D_i\Phi)^a+(\frac{g}2-1)\p(D_i\hat\Phi)^a\right)\right\}.
\end{eqnarray}
Then, it is a matter of straightforward algebra using the non-Abelian Bianchi
identity and the relation $\hat\Phi^aG_\mn^a\eq\calf_\mn-\calh_\mn$ to rewrite
this in a form analogous to (\ref{e7}),
\begin{eqnarray} \label{e15}
{\cal E}&=&\int d^3x\frac{1}2\left\{(D_i\Phi)^a+(\frac{g}2-1)\p(D_i\hat\Phi)^a
           \mp\frac{1}2\e_{ijk}\left(G_{jk}^a
           -(\frac{g}2-1)\hat\Phi^a\calh_{jk}\right)\right\}^2\no\\
      &&\pm\int d^3x\frac{1}2\e^{ijk}\pa_i\left\{\p\calf_{jk}
         -\frac{g}2\phi\calh_{jk}\right\}
        \pm\int d^3x(\frac{g}2-1)\p[\e^{ijk}\pa_i\calf_{jk}]\,.
\end{eqnarray}
The 't Hooft tensor $\calf$ satisfies its own Bianchi identity
$\e^{ijk}\pa_i\calf_{jk}\eq0$ whenever $\p\neq0$. Therefore, the last term on
the right hand side of (\ref{e15}) gives a null contribution, while the second
term may be changed to a surface integral at spatial infinity (at least for
$g>0$). Since $\calh_{jk}$ approaches zero at spatial infinity faster than
$1/r^2$, the surface integral is again given by $-4\pi n\p(\infty)/e$. Thus,
from (\ref{e15}), we have the energy bound which is independent of $g$:
\begin{equation}
{\cal E}\ge\left|\frac{4\pi n}{e}\p(\infty)\right|\,,\hspace{5mm}
\mbox{for all $g>0$}\,.
\end{equation}
On the other hand, the minimum energy configurations for any given winding
number $n$ correspond to the solutions of the generalized BPS equations which
have an explicit $g$-dependence:
\begin{equation} \label{e17}
(D_i\Phi)^a+(\frac{g}2-1)\p(D_i\hat\Phi)^a
=\pm\frac{1}2\e_{ijk}\left(G_{jk}^a-(\frac{g}2-1)\hat\Phi^a\calh_{jk}\right).
\end{equation}

As one can easily demonstrate, solutions of the above generalized BPS
equations automatically solve the full field equations. Also, if we use the
hedgehog ansatz with these equations, we easily recover (\ref{e13}). Note
that, in the unitary gauge, we can express these BPS equations by the
following first-order differential equations:
\begin{eqnarray}
\pa_i\p&=&\pm\e_{ijk}\left(\pa_jA_k-i\frac{ge}2W_j^*W_k\right),\no\\
\frac{g}2\p W_i&=&\pm\e_{ijk}\bar{D}_jW_k\,.
\end{eqnarray}
Solutions to these equations will provide us with all minimal energy magnetic
monopole solutions (for any give $n$) in the special case of the Lee-Weinberg
model, which is described by the Lagrangian density (\ref{e1}) with $V(\p)=0$,
$g^2/4=\lll$ and $m^2(\p)=\lll e^2\p^2$. But, even for $n=1$, no simple
closed-form solution is known to us yet.

Still, some comments on the unit-charged BPS solutions for arbitrary $g (>0)$
may be desirable. Note that (\ref{e13}) is invariant under a simultaneous
rescaling of $\p$ and $r$, thanks to the vanishing Higgs potential. The
resulting one-parameter family of solutions are characterized by the Higgs
expectation value $\p(\infty)$, as one would expect. Defining
$x\eq|eg\p(\infty)|r/2$, $K\eq\sqrt\frac{g}2u$ and $h\eq\p/\p(\infty)$, we
can rewrite (\ref{e13}) as
\begin{equation} \label{e19}
\frac{dK}{dx}=-hK\,,\hspace{5mm}
\frac{2}{g}\frac{dh}{dx}=\frac{1-K^2}{x^2}\,.
\end{equation}
For a finite-energy solution, we should then require that $K(\infty)=0$ and
$h(\infty)=1$. Also, near the origin, the expected behaviors for $K(x)$ and
$h(x)$ are
\begin{equation} \label{e20}
K(x)=1-K_1x^\a+\cdots\,,\hspace{5mm}
xh(x)=\a K_1x^\a+\cdots\,,
\end{equation}
where $a=\frac{1}2+\sqrt{g+\frac{1}4}$ and $K_1$ is some positive constant.
For $g=2$, the corresponding solution is well-known \cite{5}:
\begin{equation} \label{e21}
K(x)=\frac{x}{\sinh{x}}\,,\hspace{5mm}
h(x)=\frac{\cosh{x}}{\sinh{x}}-\frac{1}{x}\,.
\end{equation}
To study the case with general $g>0$, we combine the above two equations into
the following second-order equation for $L\eq-\log{K}$ (here, $s\eq\log{x}$):
\begin{equation} \label{e22}
\frac{d^2L}{ds^2}+\pa_LU_{\text{eff}}(L)=\frac{dL}{ds}\,,\hspace{4mm}
\left(U_{\text{eff}}(L)=-\frac{g}2(L+\frac{1}2e^{-2L})\right).
\end{equation}
The problem is thus changed to that of a one-dimensional particle motion in
the presence of an anti-damping force proportional to its velocity. The
condition (\ref{e20}) now reads $L(s)=K_1e^{\a s}+\cdots$ as $s\ra-\infty$,
while, from $K(\infty)=0$, we must have $L(s)\ra\infty$ as $s\ra\infty$. Now
think of $L(s)$ as the position of a ``particle'' that starts from the
``point'' $L=0$ at ``time'' $s=-\infty$. Then, once we insists $K_1$ to be
positive, the particle will always end up at $L=\infty$, because of the
continuous increase of particle ``energy'' due to the $\frac{dL}{ds}$ term in
(\ref{e22}) and the monotonically decreasing nature of $U_{\text{eff}}$ for
positive $L$. In fact, for sufficiently large $s$, we always have $L\sim
Ce^S\ra\infty$ with some positive constant $C$; then, using the first relation
in (\ref{e19}), it follows that $0<h(\infty)=C<\infty$ for generic values of
$K_1>0$. On the other hand, a particular solution of (\ref{e19}) (with the
boundary condition $h(\infty)=1$ ignored), say $\bar{h}(x)$, generates a
one-parameter family of solutions $h_\LLL(x)=\LLL\bar h(x/\LLL)$ for all
positive $\LLL$. Using this freedom, we can always find the solution that
satisfies the boundary condition $h(\infty)=1$ as desired, whenever $g>0$.
(The last step is also tantamount to choosing a particular value of $K_1$.)
This shows that there exist actual solutions saturating the BPS bound for the
unit-charged cases.

What about multi-monopole solutions? In the usual $g=2$ BPS model, static
multi-monopole solutions are possible because the repulsive electromagnetic
force is exactly balanced against the attractive scalar interaction \cite{8}.
A tell-tale sign that this cancellation occurs may be found in the spatial
components of the energy-momentum tensor $T_{ij}$ that vanishes identically
upon
using the first-order BPS equation (\ref{e9}). Remarkably, exactly the same
cancellation occurs for the present nonrenormalizable theories with arbitrary
positive $g$ also. This is because the structure of $T_{ij}$ and that of the
BPS equation remain unchanged once we replace $G_{ij}^a$ and $(D_k\Phi)^a$ by
$[G_{ij}^a-(\frac{g}2-1)\hat\Phi^a\calh_{ij}]$ and
$[(D_k\Phi)^a+(\frac{g}2-1)\p(D_k\hat\Phi)^a]$. This strongly suggests that
there should exist static multi-monopole solutions to (\ref{e17}) also. In a
way, this should have been anticipated by the following reason. As far as the
long-range interaction between monopoles are concerned, the nonrenormalizable
couplings introduced by the extra free parameter $g$ do not seem to have
significant effects. For our generalized, unit-charged BPS monopoles for
example, one can show using (\ref{e13}) that the long-distance tails are given
by
\begin{equation}
\hat\Phi^aG_{ij}^a\sim\pm\e_{ijk}\frac{\hat x^k}{er^2}\,,\hspace{4mm}
\Phi^a\sim\left[\phi(\infty)\pm\frac{1}{er}\right]\hat x^a\,,
\end{equation}
which are entirely independent of the gyromagnetic ratio $g$.

The Lee-Weinberg model allows finite-energy dyon solutions also. But a rather
surprising fact we might add here is that, if $g$ is not equal to two, our
model defined by the Lagrangian density (\ref{e2}) does {\it not} allow the
BPS-type equations for dyons. (The case of BPS dyons for $g=2$ is discussed in
Ref.\ \cite{9}.) This is partially due to the nontrivial role assumed by the
Gauss law constraint for dyon solutions.

To summarize, we have shown that a Dirac-string-free non-Abelian description
is possible for the Lee-Weinberg $U(1)$ monopoles. When the Higgs sector is
appropriately constrained, the theory admits the Bogomol'nyi-type bound which
is saturated by purely magnetic configurations solving the generalized BPS
equations. We have then explicitly demonstrated the existence of spherically
symmetric one-monopole solutions satisfying the latter equations. Mass of
these generalized BPS monopoles is independent of the gyromagnetic ratio $g$.
Also conjectured is the existence of static multi-monopole solutions in this
Bogomol'nyi limit.

It remains to see whether the generalized BPS monopoles discussed in this
paper will significantly enrich physics of solitons and lead to some new
developments in mathematical physics. In any case, it is necessary to have a
better understanding on the solution space of our generalized BPS equations.
One may also look for analogous generalized BPS systems in the context of
Yang-Mills-Higgs theories that are based on bigger gauge groups than $SO(3)$.
Another interesting issue is whether these BPS systems can be understood as
the bosonic sector of a suitable field theory possessing extended
supersymmetry \cite{10}. We hope to return to some of these outstanding
problems in near future.

\acknowledgments
P.Y. thanks K.\ Lee and E.\ Weinberg for interesting discussions, and C.L.
enjoyed useful conversations with R.\ Jackiw and H.\ Min. This work was
initiated when one of the authors (C.L.) was participating in the workshop on
``Topological Defects'' at the Isaac Newton Institute for Mathematical
Sciences in August, 1994. The work of C.L. is
supported in part by S.N.U. Daewoo Research Fund and the Korea
Science and Engineering Foundation (through the Center for Theoretical
Physics and SNU). The work of P.Y. is supported by U.S. Department of Energy.

\end{document}